\documentclass{article}

\usepackage{arxiv}
\usepackage[utf8]{inputenc} 
\usepackage[T1]{fontenc}    
\usepackage{hyperref}       
\usepackage{url}            
\usepackage{booktabs}       
\usepackage{amsfonts}       
\usepackage{nicefrac}       
\usepackage{microtype}      
\usepackage{lipsum}		
\usepackage{graphicx}
\usepackage{doi}
\usepackage{amsmath}
\usepackage{multirow}
\usepackage[square,sort,comma,numbers]{natbib}

\title{PhaGO: Protein function annotation for bacteriophages by integrating the genomic context}

\author{ \href{https://orcid.org/0009-0005-9200-4862}{\includegraphics[scale=0.06]{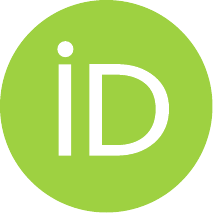}\hspace{1mm}Jiaojiao Guan} \\
	Department of Electrical Engineering\\
	City University of Hong Kong\\
	Kowloon, Hong Kong (SAR) \\
	\texttt{jiaojguan2-c@my.cityu.edu.hk} \\
	\And
	\href{https://orcid.org/0000-0001-5974-4985}{\includegraphics[scale=0.06]{orcid.pdf}\hspace{1mm}Yongxin Ji} \\
	Department of Electrical Engineering\\
	City University of Hong Kong\\
	Kowloon, Hong Kong (SAR) \\
	\texttt{yongxinji2-c@my.cityu.edu.hk} \\
 \And
	\href{https://orcid.org/0000-0002-8566-0707}{\includegraphics[scale=0.06]{orcid.pdf}\hspace{1mm}Cheng Peng} \\
	Department of Electrical Engineering\\
	City University of Hong Kong\\
	Kowloon, Hong Kong (SAR) \\
	\texttt{cpeng29-c@my.cityu.edu.hk} \\
 \And
	\href{https://orcid.org/0000-0003-2559-2519}{\includegraphics[scale=0.06]{orcid.pdf}\hspace{1mm}Wei Zou} \\
	Department of Electrical Engineering\\
	City University of Hong Kong\\
	Kowloon, Hong Kong (SAR) \\
	\texttt{weizou@cityu.edu.hk} \\
 \And
	\href{https://orcid.org/0000-0001-5974-4985}{\includegraphics[scale=0.06]{orcid.pdf}\hspace{1mm}Xubo Tang} \\
	Department of Electrical Engineering\\
	City University of Hong Kong\\
	Kowloon, Hong Kong (SAR) \\
	\texttt{xubotang2-c@my.cityu.edu.hk} \\
 \And
	\href{https://orcid.org/0000-0001-5974-4985}{\includegraphics[scale=0.06]{orcid.pdf}\hspace{1mm}Jiayu Shang} \thanks{Co-corresponding author.}\\
	Department of Electrical Engineering\\
	Chinese University of Hong Kong\\
	Kowloon, Hong Kong (SAR) \\
	\texttt{jiayushang@cuhk.edu.hk} \\
  \And
	\href{https://orcid.org/0000-0003-1373-8023}{\includegraphics[scale=0.06]{orcid.pdf}\hspace{1mm}Yanni Sun}\footnotemark[1]\\
	Department of Electrical Engineering\\
	City University of Hong Kong\\
	Kowloon, Hong Kong (SAR) \\
	\texttt{yannisun@cityu.edu.hk} 
}

\renewcommand{\shorttitle}{\textit PhaGO}

\hypersetup{
pdftitle={PhaGO: Protein function annotation for bacteriophages by integrating the genomic context},
pdfsubject={q-bio.NC, q-bio.QM},
pdfauthor={Jiaojiao Guan,Yongxin Ji,Cheng Peng,Wei Zou,Xubo Tang,Jiayu Shang,Yanni Sun},
pdfkeywords={Bacteriophage, Protein function annotation},
}

\begin{document}
\maketitle

\begin{abstract}
Bacteriophages are viruses that target bacteria, playing a crucial role in microbial ecology. Phage proteins are important in understanding phage biology, such as virus infection, replication, and evolution. Although a large number of new phages have been identified via metagenomic sequencing, many of them have limited protein function annotation. Accurate function annotation of phage proteins presents several challenges, including their inherent diversity and the scarcity of annotated ones. Existing tools have yet to fully leverage the unique properties of phages in annotating protein functions. In this work, we propose a new protein function annotation tool for phages by leveraging the modular genomic structure of phage genomes. By employing embeddings from the latest protein foundation models and Transformer to capture contextual information between proteins in phage genomes, PhaGO surpasses state-of-the-art methods in annotating diverged proteins and proteins with uncommon functions by 6.78\% and 13.05\% improvement, respectively. PhaGO can annotate proteins lacking homology search results, which is critical for characterizing the rapidly accumulating phage genomes. We demonstrate the utility of PhaGO by identifying 688 potential holins in phages, which exhibit high structural conservation with known holins. The results show the potential of PhaGO to extend our understanding of newly discovered phages.
\end{abstract}


\section{Introduction}
Bacteriophages (phages) are viruses that can infect bacterial cells. They are highly prevalent and abundant in the biosphere, being found in various environmental matrices, including gastrointestinal tracts of animals, water bodies, and soil \citep{cobian2016viruses, zeng2024metagenomic, WANG2024120859}. Accumulating studies have demonstrated the important role of phages in microbial communities. For example, phages have been observed to facilitate the horizontal transfer of genes between bacteria, which can influence bacterial adaptation, evolution, and acquisition of new functionalities \citep{fernandez2018phage}. In addition, they can modulate the abundance and diversity of bacterial populations by killing their host \citep{diaz2014bacteria}. Due to the increasing threats posed by antibiotic resistance, phages have gained significant attention as potential alternatives to traditional antibiotics, as they can lyse pathogenic bacteria. \citep{Bacteriophage_therapy, Bacteriophage_therapy2, azimi2019phage}.

Despite the significance of phages, the efficacy of their applications heavily relies on prior knowledge of protein functions. Understanding the protein function enables us to identify phage proteins that can target and disrupt essential bacterial processes, offering the potential for the development of targeted antimicrobial therapies \citep{shibayama2011phage}. For example, holin proteins, known for their cell-killing capabilities and broad host range, have gained significant attention for their potential applications in bacterial control. \citep{santos2018exploiting, song2016identification}. To accelerate the application of phages, it is crucial to figure out the annotation of the proteins in phages.


Gene Ontology (GO) terms are widely used to annotate the phage proteins. They are standardized vocabulary and hierarchical frameworks comprising three key dimensions: biological process (BP), cellular component (CC), and molecular function (MF) \citep{gene2019gene}. BP encompasses the sequences of events or pathways in which proteins participate, such as cellular signaling or metabolic processes, while CC pertains to the subcellular locations or structures where proteins are localized, such as the nucleus or plasma membrane. The MF aspect centers on the distinct activities and tasks carried out by proteins, such as enzyme catalysis or receptor binding.  

However, there are two major challenges to using GO terms to annotate phage protein. First, the number of phage proteins with known GO labels is limited. Until February 27, 2024, the total number of phage proteins from the National Center for Biotechnology Information Reference Sequence Database (NCBI RefSeq) is $541,060$, derived from $5,160$ complete genomes. However, only 20.85\% percent of proteins have GO labels. This scarcity of labeled proteins results in an insufficient database for comprehensive functional annotation. Second, although phage encodes a small number of proteins compared to their hosts, these proteins exhibit a remarkable degree of functional diversity. For example, among the 1173 phage proteins provided by the UniProtKB database, there are a total of 912 GO terms. This means that on average each GO label contains less than two supporting samples and will bring challenges to computational methods. Moreover, the distribution of these GO terms is imbalanced, with certain terms being more prevalent or specific compared to others. This imbalanced label distribution poses a significant challenge to accurate classification. These obstacles impose great requirements on annotation tools.

Several attempts have been made to analyze and annotate protein functions. They can be categorized into two types: homology-based and deep learning-based methods. The summarized information of the state-of-art methods is listed in Table \ref{tab:all_methods_introduction}. Homology-based methods, such as DiamondScore \citep{DeepGOplus} and DiamondBlast \citep{DeepGOplus}, rely on sequence similarity to infer protein function. These methods assume that proteins with similar sequences share similar functions. However, due to the extensive genetic diversity and rapid evolution of phages, phage proteins may not have a significant sequence similarity when aligned to the reference database.

In order to annotate more proteins, most deep-learning methods formulate protein function annotation as a multi-label prediction task, where protein sequences or extracted features are used as the model input, and the predicted GO terms represent outputs. For example, DeepGOPlus utilizes sequence information alone and employs convolutional neural networks (CNN) to scan the protein sequence for motifs \citep{DeepGOplus}. Then, it combines the sequence-based predictions with an alignment-based search. ATGO \citep{zhu2022integrating} utilized the ESM-1b \citep{rives2021biological}, a pre-trained large language model, to extract embeddings of protein sequences from the last three layers. Through a triplet network, ATGO learns embeddings that enhance the proximity of proteins with similar functions in the embedding space while promoting greater separation among proteins with different functions. In contrast to numerous methods that disregard the correlations of GO labels, PFresGO \citep{pan2023pfresgo} integrates the hierarchical inter-relationship of GO using Anc2Vec \citep{edera2022anc2vec}. A pre-trained large language model called ProtT5 is utilized to extract the protein embedding. Furthermore, PFresGO employs a cross-attention mechanism to effectively capture the connections between local protein residues and GO terms, enhancing the accuracy and specificity of protein annotation. However, the methods described previously overlook the unique properties specific to phage proteins. Thus, there is considerable potential for enhancing the annotation of phage proteins. In our investigation, we have discovered that the order of phage protein functions exhibits a high level of conservation within the same genus. It means that the surrounding context proteins can provide valuable insights for predicting protein functions in phages.

In this work, we present a novel method, PhaGO, for phage protein annotation by integrating the powerful foundation model with the unique properties of phages. There are two main steps in our PhaGO framework. First, we utilize a pre-trained protein language model (PLM), ESM2 \citep{esm2}, to encode phage proteins. ESM2 has acquired a comprehensive understanding of various protein features, including aspects such as three-dimensional structure and interaction relationships during training. Thus, it can effectively return meaningful representations for phage proteins. Second, we reformat the phage genomes into protein sentences using embeddings obtained from the PLM. Then, we train a Transformer-based natural language model to learn and leverage inherent order association among phage proteins. By considering the positions of proteins and their functions within the genomic context, the model is expected to achieve further improvements in phage protein annotation. The experiments demonstrated a significant advantage of PhaGO in accurately predicting GO terms, achieving impressive AUPR scores of 0.8636, 0.8882, and 0.8277 for BP, CC, and MF ontology, respectively. Notably, PhaGO showcased substantial improvements in predicting the functions of proteins that lacked alignment with the database and minority GO labels, addressing an important challenge in functional annotation. In the case study, PhaGO demonstrates great promise in unraveling the functions of key phage proteins that lack alignment with the reference database. We identified 688 holin proteins and showed prediction reliability based on structural homology. Thus, PhaGO has the potential to accelerate and enhance the comprehensive understanding of phages and their biological processes.
\begin{table}[t]
\centering
\caption{The introduction of recent protein function annotation tools. `DL' means the deep learning-based tool.}
\setlength{\tabcolsep}{1 mm}{
\begin{tabular}{lll}
\hline
Tool         & Type          & Method             
\\
\hline
DiamondScore    & Homology      & DIAMOND BLASTP \\
DiamondBlast & Homology      & BLAST          \\
DeepGOCNN     & DL & CNN                                      \\
DeepGOPlus    & Hybrid        & CNN + DIAMOND BLASTP                       \\
ATGO     & DL & ESM-1b + Triplet neural network        \\
PFresGO   & DL  & ProtT5 + GO term relationship 
\\
\hline
\end{tabular}
}
\label{tab:all_methods_introduction}
\end{table}

\begin{figure}[b] 
  \centering
  \includegraphics[width=\linewidth]{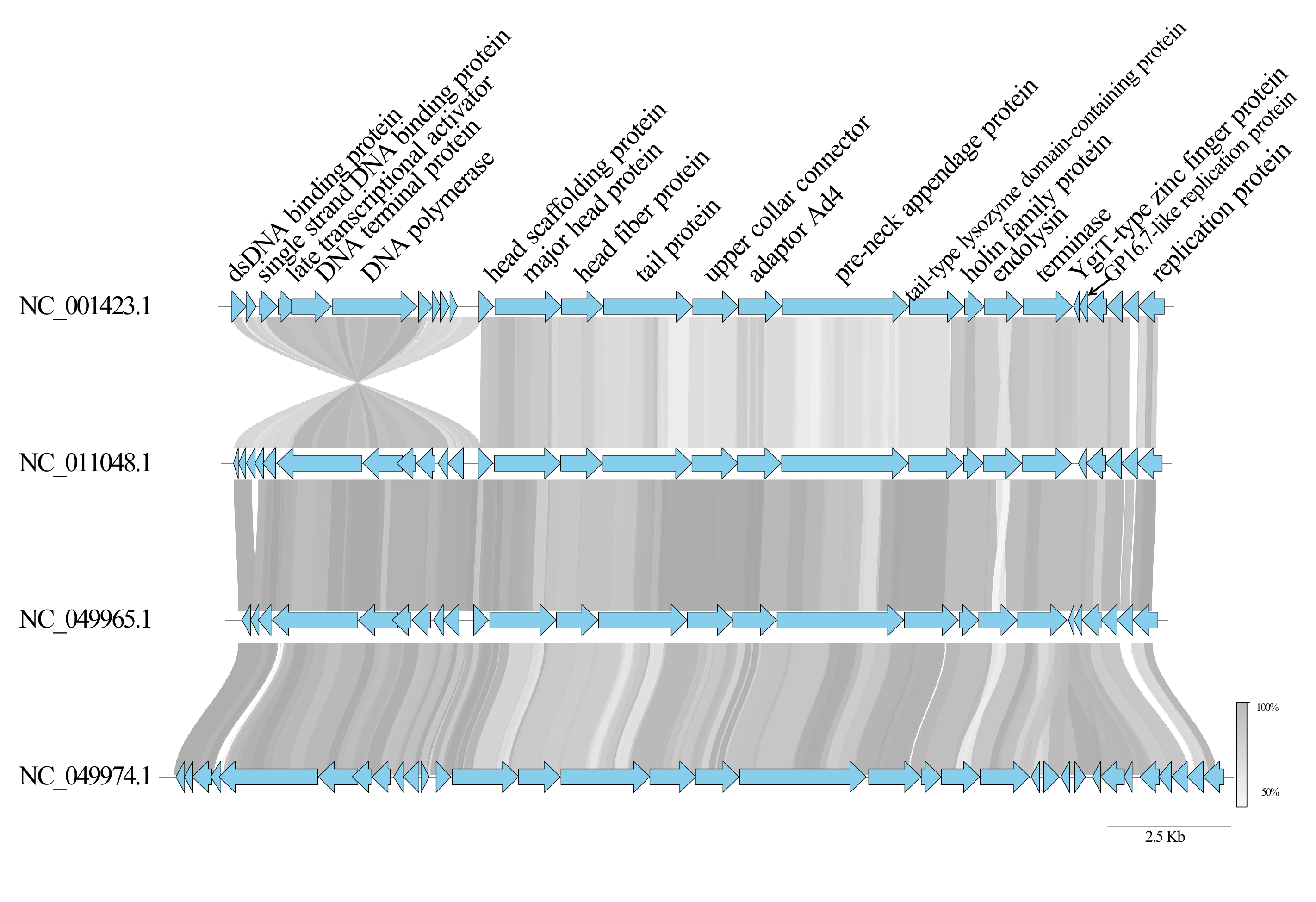}
  \caption{The function order of proteins within four phage genomes. The blue arrow represents the protein and the grey link shows the similarity among proteins.}
 \label{fig:genus_order}
\end{figure}

\section{Methods and materials}
\label{sec:headings}

The proteins in the phage sequences are similar to the words in the natural language. Thus, the phage genomes can be viewed as a language of phage life that exhibits distinct features. One notable observation of these phage languages is that phage proteins within the same genus tend to maintain a consistent arrangement. For instance, Fig. \ref{fig:genus_order} reveals a distinct pattern in the order of proteins within the \textit{Salasvirus} genus. These characteristics inspire us to reformat the phage genomes into sentences with contextual proteins and predict the annotations based on the surrounding information. In the following section, we will detail how PhaGO leverages the contextual information for phage protein annotation.


\begin{figure*}[t] 
  \centering
  \includegraphics[width=\linewidth]{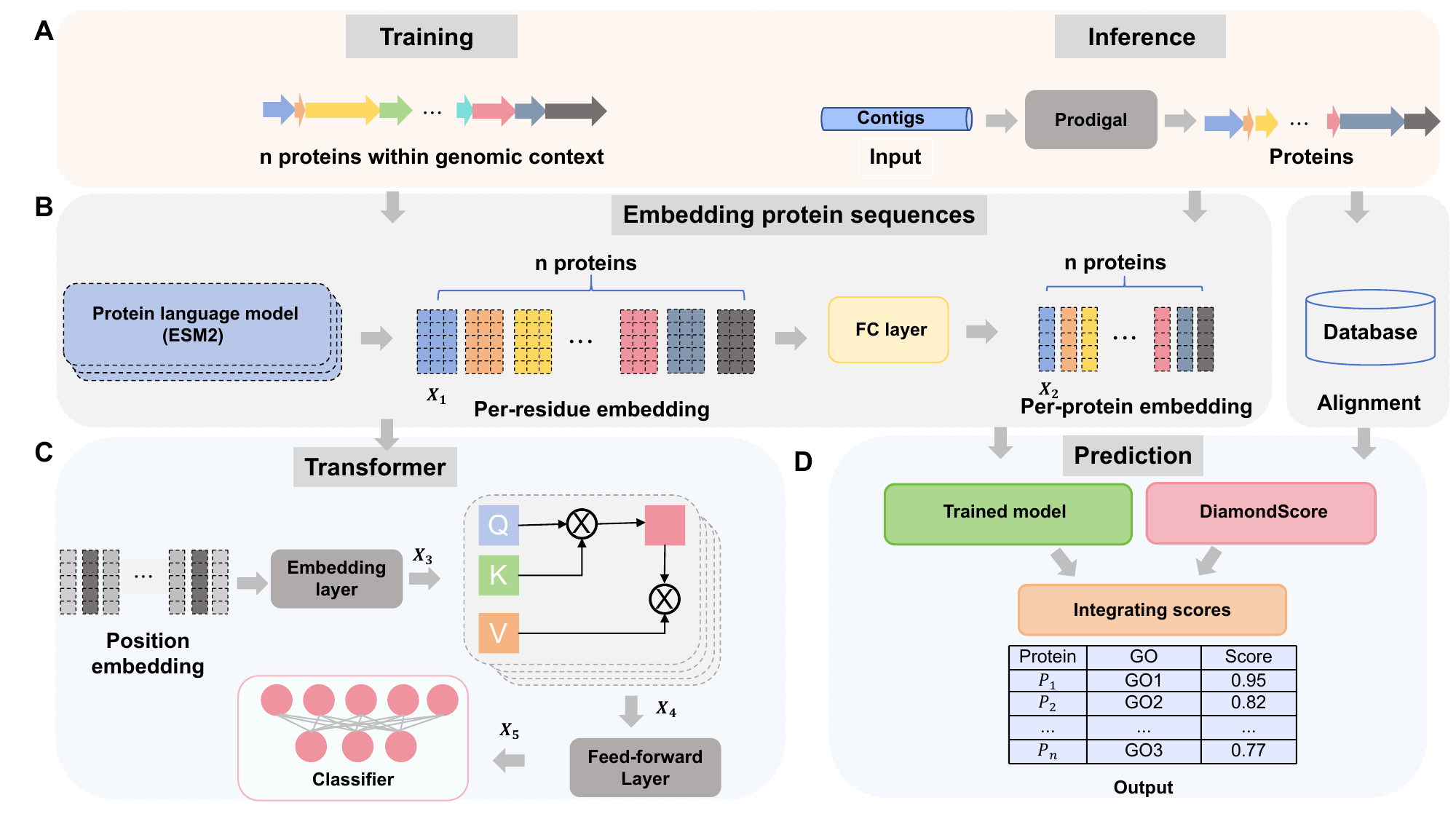}
  \caption{The architecture of PhaGO for phage protein function prediction. Data processing steps for training and inference are depicted in (A). Then in (B), the ESM2 model is employed to extract per-residue embeddings, followed by the utilization of a fully connected (FC) layer to generate a compact protein embedding. The Transformer model is utilized to learn the contextual relationship in (C). Combined with the alignment-based method, we obtain the GO terms prediction score in (D).}
 \label{fig:architecture}
\end{figure*}

\subsection{Embedding protein sequences}
Fig. \ref{fig:architecture} shows the architecture of the PhaGO model. In Fig. \ref{fig:architecture}A, let the number of proteins of a phage genome in the training process be $n$. The first step is to encode the phage genomes by generating the embedding of the $n$ proteins. To obtain protein embedding, we employ the ESM2 model which is pre-trained on protein sequences sourced from UR50/D. During training, ESM2 selects 15\% amino acids for masking and predicts amino acids at the masked position. Based on a third-party benchmark result \citep{outeiral2024codon}, ESM2-33 has a better performance than the ProtT5 family. Moreover, the performance of the ESM2-33 is comparable with ESM2-36 and ESM2-48, but the latter two models have more parameters, leading to a significant increase in runtime. Specifically, the ESM2-33 model consists of approximately 650 million parameters, while the ESM2-36 and ESM2-48 models contain 3 billion and 15 billion parameters, respectively. Therefore, we chose ESM2-33 to embed the proteins. 


We define $d_e$ as the dimension of per-residue embedding and impose a maximum limit of 1,024 residues for each protein sequence, which aligns with the default setting of the ESM2 model. By applying the ESM2 embedding to the protein sequences and considering these constraints, we generate an embedding matrix $X_1$ with dimensions of $1,024 \times d_e$ for each protein shown in Fig. \ref{fig:architecture}B. In the ESM2-33 model, the default value of $d_e$ is 1,280. Subsequently, we pass $X_1$ through a fully connected (FC) layer, resulting in a one-dimensional feature set denoted as $X_2$.





\subsection{Learning the relationship of context proteins using Transformer}

As words and sentences in human language derive meaning through their context and relationships with other linguistic elements, proteins can also be better understood by considering their interactions, dependencies, and roles within the genome. Therefore, we annotate the protein functions by considering the context neighbors. This goal is achieved through two steps: preparing the context protein embedding and learning the relationships among proteins within the same genome. The sequential steps are illustrated in Fig. \ref{fig:architecture}C.

To obtain the context protein embedding, first, we treat each protein as a token and contigs can be seen as sentences composed of multiple tokens. Then, we combine the embeddings of each protein into a single embedding with dimensions of $n \times d_e$. This integration process takes into account the order in which the proteins appear in the contigs and allows us to preserve the contextual relationships among the proteins within the same genomic context.

To incorporate positional information, we utilize position embedding. The position embedding component takes the index of each protein as its input and generates an embedding vector that encodes the relative position of the token within the sequence. This allows the model to understand the relative distance of the proteins. The final output $X_3$ of the embedding layer is obtained by summing the context protein embedding and position embedding results, resulting in a comprehensive representation of each protein in the sequence.

After embedding the context proteins into an $n \times d_e$ matrix, we introduce a crucial component in our architecture: the self-attention layer. This layer plays a vital role in learning intricate connections between proteins. To perform self-attention computations in Eqn. \ref{Eq:self_attention}, we transform the input matrix into three separate matrices: Query (Q), Key (K), and Value (V) through three independent FC layers. The $n \times n$ attention matrix is computed by multiplying the Q and K, representing the strength of protein associations. To prevent excessive values, we scale the attention matrix by dividing it by the square root of the dimension of matrix K (denoted as $\sqrt{d_k}$). Next, we normalize the attention matrix using the softmax function, assigning weights to protein pairs to indicate their relative importance. Finally, we score the proteins in the sequence by multiplying the V with the weight matrix. 

In order to collectively focus on information stemming from diverse representation subspaces, we employ a multi-head mechanism in Eqn. \ref{Eq:multihead_attention}, where each head represents a separate self-attention layer. Computation is performed in parallel across all heads, and then the concatenated head is input into an FC layer shown in Eqn. \ref{Eq:contact}, $W^M \in \mathbb{R}^{1280 \times 1280}$. Following the multi-head attention block, the resulting output $X_4$ is subsequently passed through a feed-forward layer.

\begin{equation}
\text { Attention }(Q, K, V)=\operatorname{softmax}\left(\frac{Q K^T}{\sqrt{d_k}}\right) V
\label{Eq:self_attention}
\end{equation}

\begin{equation}
\text {head}_i=\text { Attention }\left(Q_i, K_i, V_i\right), 
\label{Eq:multihead_attention}
\end{equation}

\begin{equation}
X_4=FC( \operatorname{Contact}\left(\text {head}_1, \ldots, \text {head}_n\right), W^M),
\label{Eq:contact}
\end{equation}

\subsection{Predicting the GO terms}

We formulate the protein function annotation task as a multi-label binary classification task. The goal is to assign a probability to each GO term, indicating the likelihood of the protein being associated with that specific function. The results of the feed-forward layer $X_5$ are input into a fully connected layer with the sigmoid activation function and the output is an m-dimensional vector, where $m$ represents the number of GO terms shown in Eqn. \ref{classification_equation}. 

\begin{equation}
Y=\operatorname{sigmoid}\left(W \cdot  X_5+b\right) .
\label{classification_equation}
\end{equation}

During training, the model is optimized by minimizing the binary cross-entropy loss. This loss function in Eqn. \ref{loss_equation} is commonly used in binary classification tasks to measure the difference between predicted probabilities and actual labels. In addition, we train three GO prediction models on BP, CC, and MF separately.

\begin{equation}
\mathcal{L}=-\frac{1}{N} \sum_{i=1}^N \sum_{j=1}^{|\mathrm{GO}|} y_{i j} \log \left(\hat{y}_{i j}\right)
\label{loss_equation}
\end{equation}

\subsection{Integrating PhaGO with alignment-based method}

Considering that proteins with significant alignment usually have high-precision GO prediction results\citep{DeepGOplus, cao2021tale}, we introduce a hybrid mode named $\mathrm{PhaGO}^+$ by incorporating DiamondScore into the PhaGO to enhance the predictive capabilities for phage protein annotations.

\begin{equation}
\mathrm{S}_{\mathrm{PhaGO}^+}(i) = \beta \cdot \mathrm{S}_{\mathrm{PhaGO}}(i)+(1-\beta) \mathrm{S}_{\mathrm{DiamondScore}}(i)
\end{equation}

where $\mathrm{S}_{\mathrm{PhaGO}^+}(i)$ is the confidence score of $\mathrm{PhaGO}^+$ for protein $i$, $\mathrm{S}_{\mathrm{PhaGO}}(i)$ and $\mathrm{S}_{\mathrm{DiamondScore}}(i)$ are confidence scores of PhaGO and DiamondScore. The values of the weight parameter $\beta$ are fine-tuned based on the validation dataset.

\section{Results}
\label{sec:others}

\subsection{Metrics}
We evaluate the performance of PhaGO following previous work. Specifically, we present two sets of metrics, corresponding to the prediction accuracy of protein-centric and GO term-centric evaluation, which are used in the Critical Assessment of Functional Annotation (CAFA) competitions. The protein-centric evaluation focuses on determining the function prediction accuracy, whereas the term-centric evaluation aims to examine whether the model can correctly identify proteins associated with a particular functional term \citep{radivojac2013large}. The latter can provide the performance of different function terms. 



First, we introduce protein-centric metrics. Let $P_i(t)$ be the set of GO terms for a protein $i$ returned by the model under the score cutoff $t$, while $T_i$ represents the true GO term set for protein $i$. Then recall and precision for each protein $i$ with threshold $t$ are calculated in Eqn. \ref{Eq:pro_cen_recall} and Eqn. \ref{Eq:pro_cen_precision}. To calculate the average recall and precision on all proteins, we define $n$ as the total number of proteins and $n_t$ as the number of proteins that have at least one predicted GO term when the threshold is $t$. The equations are shown in Eqn. \ref{Eq:average_recall} and Eqn. \ref{Eq:average_precision}, respectively. We record the F1-score calculated for each threshold $t$, ranging from 0 to 1, and obtain the maximum F1-score as $F_{\max}$ shown in Eqn. \ref{Eq:pro_cen_fmax}. To compute the Area Under the Precision-Recall Curve (AUPR), the prediction scores of proteins are concatenated and input into the \textit{scikit-learn} Python package.

\begin{equation}
recall_i(t)=\frac{ | P_i(t) \cap T_i |}{|T_i|}
\label{Eq:pro_cen_recall}
\end{equation}

\begin{equation}
pre_i(t)=\frac{|P_i(t) \cap T_i|}{|P_i(t)|}
\label{Eq:pro_cen_precision}
\end{equation}

\begin{equation}
AvgRecall(t)=\frac{1}{n} \cdot \sum_{i=1}^n recall_i(t)
\label{Eq:average_recall}
\end{equation}

\begin{equation}
AvgPre(t)=\frac{1}{n_t} \cdot \sum_{i=1}^{m(t)} p re_i(t)
\label{Eq:average_precision}
\end{equation}

\begin{equation}
F_{\max }=\max _t\left\{\frac{2 \cdot \operatorname{AvgPre}(t) \cdot \operatorname{AvgRecall}(t)}{\operatorname{AvgPre}(t)+\operatorname{AvgRecall}(t)}\right\}
\label{Eq:pro_cen_fmax}
\end{equation}


Then, we present the term-centric evaluation. To calculate the term-centric $F_{\max}$, we follow a three-step process. First, we calculate the precision and recall for GO term $l$ under threshold $t$, as defined in Eqn. \ref{Eq:term_precision} and \ref{Eq:term_recall}. In the second step, we calculate $F_{\max}(l)$, which is the maximum F1-score for label $l$ under different score cutoffs (Eqn. \ref{Eq:term_fmax}). Finally, we average these $F_{\max}(l)$ values across all GO labels to obtain the final $F_{\max}$, as shown in Eqn. \ref{Eq:final_fmax}. The AUPR for each label is calculated and averaged to obtain the final AUPR. 



\begin{equation}
pre_l(t)=\frac{\sum_i I\left(l \in P_i(t) \wedge l \in T_i\right)}{\sum_i I\left(l \in P_i(t)\right)}
\label{Eq:term_precision}
\end{equation}

\begin{equation}
recall_l(t)=\frac{\sum_i I\left(l \in P_i(t) \wedge l \in T_i\right)}{\sum_i I\left(l \in T_i\right)}
\label{Eq:term_recall}
\end{equation}

\begin{equation}
F_{\max } (l)=\max _t\left\{\frac{2 \cdot pre_l(t) \cdot recall_l(t)}{p re_l(t)+recall_l(t)}\right\}
\label{Eq:term_fmax}
\end{equation}

\begin{equation}
F_{\text {max }}=\sum_{l=0}^m F_{\max }(l)
\label{Eq:final_fmax}
\end{equation}

\subsection{Dataset}

We downloaded the reference genomes and proteins under the Caudoviricetes class from the NCBI RefSeq database. Due to the lack of GO terms in the Refseq database, we mapped the protein accessions into UniProt database \citep{uniprot2023uniprot} using the `ID mapping' tool and retrieved annotations. 

\begin{table*}[b]
\centering
\caption{Performance comparison of PhaGO/$\mathrm{PhaGO}^+$ and state-of-the-art methods for protein function prediction based on term-centric evaluation. $\mathrm{PhaGO}_{\mathrm{BASE}}$ and $\mathrm{PhaGO}_{\mathrm{LARGE}}$ are based on ESM2-12 and ESM2-33 foundation model, respectively. `+' means the methods are integrated with the alignment-based tools.}
\setlength{\tabcolsep}{6 mm}{

\begin{tabular}{lcccccc}
\hline
                   & \multicolumn{2}{c}{BP}                                          & \multicolumn{2}{c}{CC}                                          & \multicolumn{2}{c}{MF}                                          \\
\hline
\multirow{-2}{*}{} & \multicolumn{1}{l}{AUPR}       & \multicolumn{1}{l}{Fmax}       & \multicolumn{1}{l}{AUPR}       & \multicolumn{1}{l}{Fmax}       & \multicolumn{1}{l}{AUPR}       & \multicolumn{1}{l}{Fmax}       \\
\hline
DiamondScore       & 0.7225                         & 0.6710                         & 0.7552                         & 0.6269                         & 0.6557                         & 0.6446                         \\
DeepGOCNN             & 0.6222                         & 0.6380                         & 0.6353                         & 0.6455                         & 0.4348                         & 0.4590                         \\
DeepGOPlus         & 0.7279                         & 0.7349                         & 0.7623                         & 0.7489                         & 0.6304                         & 0.6590                         \\
PFresGo            & 0.7642                         & 0.7692                         & 0.8232                         & 0.8026                         & 0.7210                         & 0.7430                         \\
$\mathrm{PhaGO}_{\mathrm{BASE}}$      & 0.7946                         & 0.7814                         & 0.8636                         & 0.8108                         & 0.7368                         & 0.7505                         \\
$\mathrm{PhaGO}_{\mathrm{LARGE}}$       & 0.8382                         & 0.8115                         & 0.8664                         & 0.8399                         & 0.8125                         & 0.7974                         \\
$\mathrm{PhaGO}^+_{\mathrm{BASE}}$      & 0.8595                         & 0.8263                         & \textbf{0.8882} & 0.8410                          & 0.7804                         & 0.7870                         \\
$\mathrm{PhaGO}^+_{\mathrm{LARGE}}$     & \textbf{0.8636} & \textbf{0.8341} & 0.8783                         &  \textbf{0.8493} & \textbf{0.8277} & \textbf{0.8095} \\

\hline
\end{tabular}

}
\label{tab:all_methods_term_centric}
\end{table*}


To ensure an adequate number of labeled proteins for training, we labeled the proteins with no GO terms using the Prokaryotic Virus Remote Homologous Groups (PHROG) database \citep{terzian2021phrog} based on HHsuite tool \citep{steinegger2019hh}. The database contains 38,880 PHROGs, which encompass a total of 868,340 proteins derived from complete genomes of viruses infecting bacteria or archaea. Moreover, we saved the hits that demonstrated a probability of the template being homologous to query sequences exceeding 80\%, ensuring the reliability and high confidence of the matches between the phage proteins and the entries in the PHROG database. Although we used the pairwise alignment to extend the dataset, the proteins with significant alignments were only 15.51\%. The remained 63.64\% proteins still lacked annotations, which further demonstrated the necessity and importance of developing an effective phage protein annotation tool. 

Because of the requirement for rich contextual information for proteins, we selectively focused on proteins from genera with high annotation rates. The annotation rate for each genome is calculated below.

\begin{equation}
\text { Annotation Rate }=\frac{\text { \#proteins with annotation }}{\text { \#total proteins }}
\end{equation}

Then, we computed the average annotation rate of the complete genomes in each genus. Proteins from genera where the annotation rates exceeded 30\% for the BP and MF categories and 20\% for the CC category are included. It was important to note that the number of proteins annotated by CC terms was relatively smaller compared to BP and MF. Therefore, we set a lower threshold for CC to ensure including more genera. By setting these thresholds, we aimed to focus on genera with more comprehensive annotation information. The annotation rates for all genera can be found in the Supplementary material.

Additionally, we excluded single-genome genera from our considerations because they were inadequate for training. Consequently, 62 genera, 59 genera, and 203 genera were retained for BP, CC, and MF. For our analysis, we utilized the proteins exclusively from these selected genera. we aim to minimize the similarity between the training and test/validation datasets during the dataset splitting process. Thus, we implement the following steps for dataset partitioning:

\begin{itemize}

\item The proteins obtained from the selected genera are aligned all against all using the DIAMOND BLASTP \citep{buchfink2015fast} with a default e-value threshold of 0.001. The alignment scores among proteins are used to build a graph. Then Markov clustering algorithm (MCL) \citep{mcl} is applied to cluster the protein graph, which is a fast and unsupervised method.

\item We randomly select clusters and include all proteins within those clusters in the training dataset until the cumulative size exceeds 80\% of the total dataset. The remaining proteins are placed in an independent dataset for evaluation purposes.

\item Finally, we randomly divide the independent dataset into two equal-sized parts while ensuring an even distribution of proteins for each GO term label.

\end{itemize}

The GO labels are obtained by propagating all ancestors based on the `is\_a' relationship in the tree. Then, we calculate the number of proteins annotated by each GO term and filter out terms with fewer than 200 annotated proteins in the training dataset. We follow the standard practice of CAFA assessment and exclude the root terms. The final protein and label number for three ontologies are shown in Supplementary Table 1.

To enhance the user experience, we provide two variant versions of PhaGO. The first version, named $\mathrm{PhaGO}_{\mathrm{LARGE}}$, is based on ESM2-33 and offers superior performance at the cost of increased computational resources and runtime. The second variant $\mathrm{PhaGO}_{\mathrm{BASE}}$ utilizes ESM2-12, providing a lightweight alternative with reduced computational demands. To be specific, we conducted tests on the prediction runtime for 1000 proteins. The results indicate that $\mathrm{PhaGO}_{\mathrm{LARGE}}$  takes 7.74 minutes, which is around 3.5 times longer than the runtime of $\mathrm{PhaGO}_{\mathrm{BASE}}$. Moreover, the parameter count for $\mathrm{PhaGO}_{\mathrm{LARGE}}$ is approximately 7 times higher than that of $\mathrm{PhaGO}_{\mathrm{BASE}}$, as outlined in Supplementary Table 2. 



\subsection{PhaGO outperforms the state-of-the-art predictors}

In this experiment, we compared PhaGO with four tools: DiamondScore\cite{DeepGOplus}, DeepGOCNN\cite{DeepGOplus}, DeepGOPlus\cite{DeepGOplus}, and PFresGO\cite{pan2023pfresgo}. These tools are the most widely used pipelines for general protein function annotation and have been demonstrated as state-of-the-art predictors. The same training dataset was utilized for retaining the learning-based methods (DeePGOCNN, DeepGOPlus, and PFresGO) or constructing the database for the alignment-based methods (DiamondScore). The performance evaluation was then carried out using the same test dataset, which ensured a fair and comparable assessment for all methods.

The performance based on term-centric is presented in Table. \ref{tab:all_methods_term_centric}, while the results obtained from protein-centric evaluation can be found in Supplementary Table 3. $\mathrm{PhaGO}^+$ outperforms the second-best method, regarding both AUPR and Fmax scores with notable improvements across all three categories. Specifically, the improvements of 9.94\%, 6.50\%, and 10.67\% in AUPR  and 6.49\%, 4.67\%, and 6.65\% in Fmax scores for BP, CC, and MF, respectively.


Comparing $\mathrm{PhaGO}_{\mathrm{BASE}}$ and $\mathrm{PhaGO}_{\mathrm{LARGE}}$, the results reveal that using a larger protein foundation model has a better performance because of the larger foundation model's ability to capture and learn more intricate biological signals. The most significant improvement is observed in the MF category, with a notable increase of 7.57\% in AUPR and 4.69\% in Fmax.

Additionally, integrating DiamondScore with PhaGO through hybrid approaches can further improve the performance in protein function prediction. Comparing $\mathrm{PhaGO}$ and $\mathrm{PhaGO}^+$, the BP category exhibits a highest improvement of 6.49\% and 4.49\% in AUPR and Fmax for $\mathrm{PhaGO}^+_{\mathrm{BASE}}$  and 2.54\% and 2.26\% in AUPR and Fmax for $\mathrm{PhaGO}^+_{\mathrm{LARGE}}$. 

Taken together, utilizing a deeper foundation model and integrating homologous search methods can help PhaGO achieve the best performance in protein function prediction. In addition, based on the performance, $\mathrm{PhaGO}^+_{\mathrm{BASE}}$ and $\mathrm{PhaGO}^+_{\mathrm{LARGE}}$ are recommended for users. However, in scenarios where computational resources are constrained, $\mathrm{PhaGO}^+_{\mathrm{BASE}}$ is the preferable choice.

\begin{figure*}[t] 
  \centering
\includegraphics[width=\linewidth]{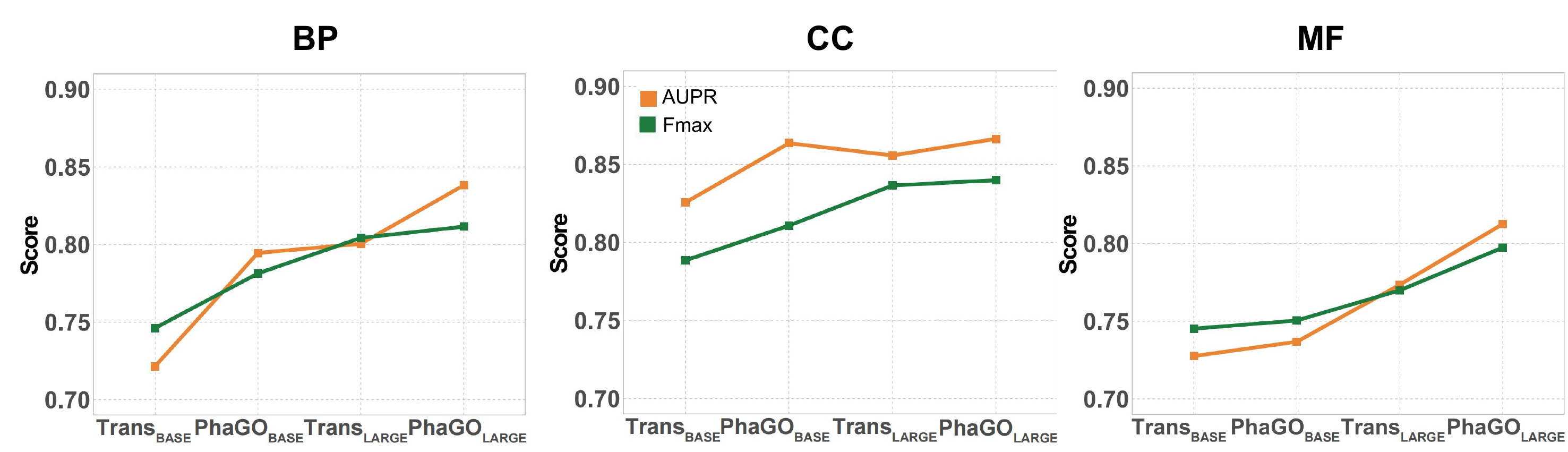}
  \caption{The effect of the context proteins for performance based on AUPR and Fmax of term-centric in three ontologies. }
 \label{fig:context}
\end{figure*}

\begin{figure}[b] 
  \centering
\includegraphics[width=\linewidth]{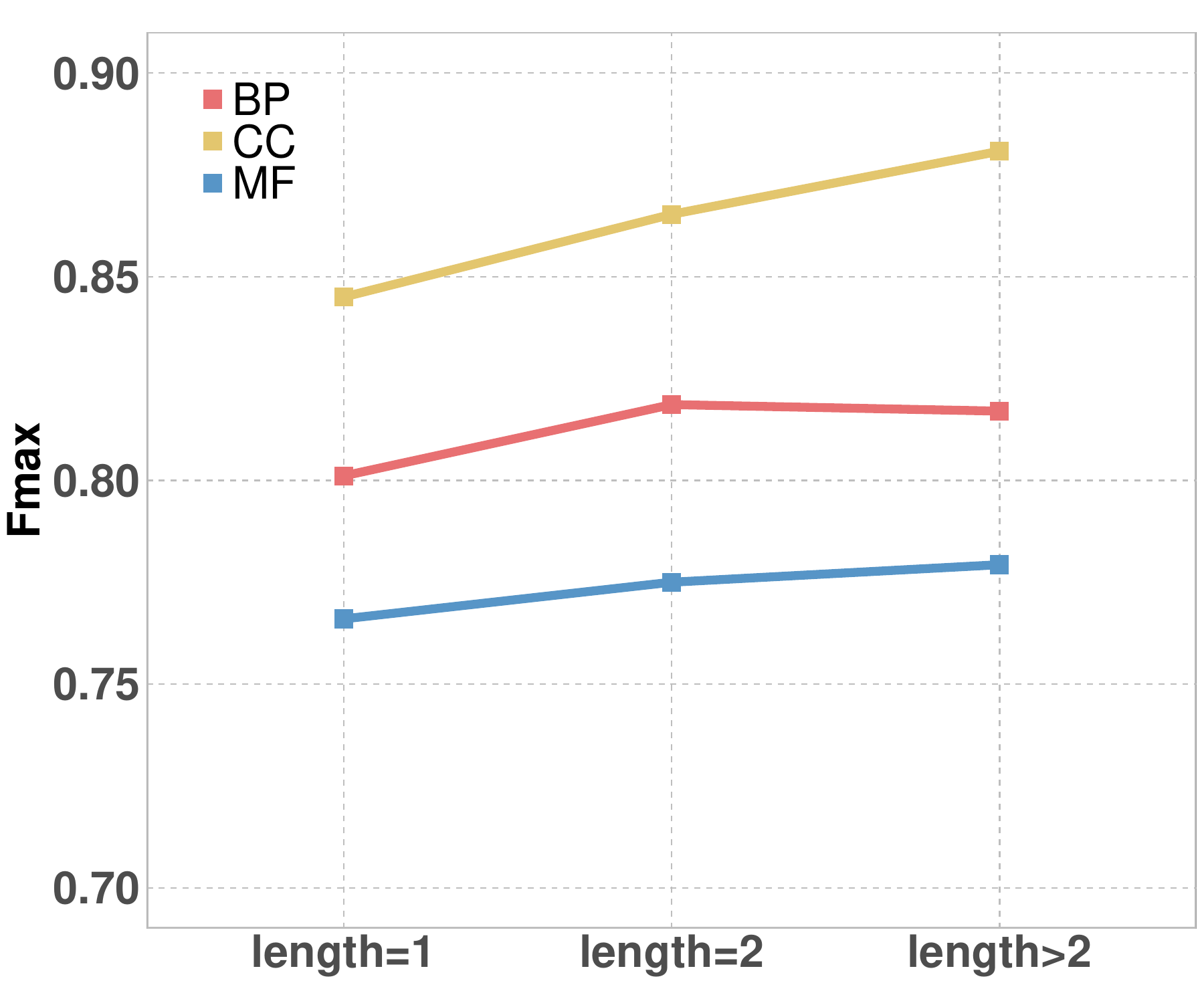}
  \caption{The impact of increasing the number of context proteins on performance in protein function prediction using Fmax of protein-centric.}
 \label{fig:context_length}
\end{figure}

\begin{figure*}[t] 
  \centering
\includegraphics[width=\linewidth]{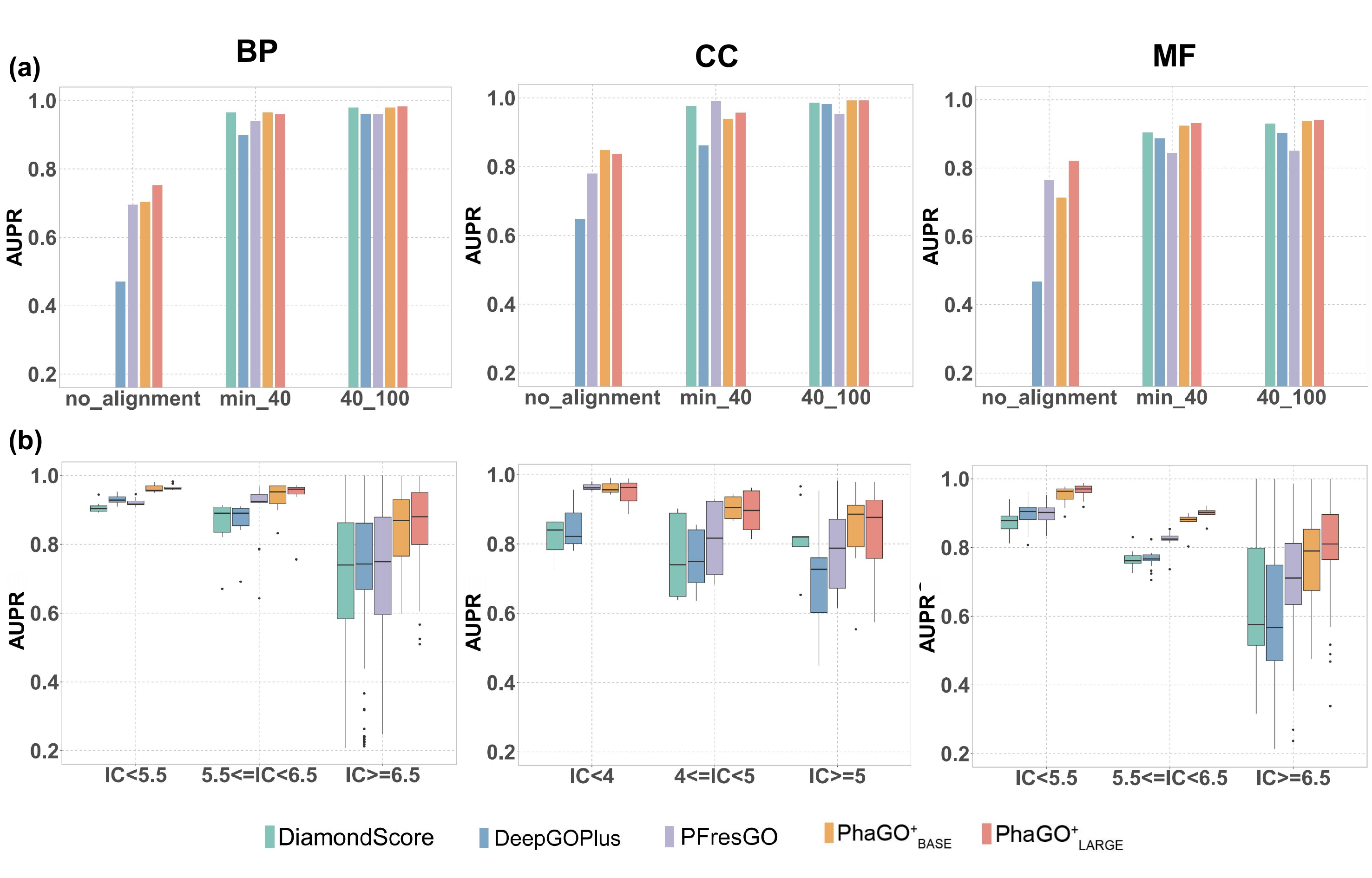}
  \caption{Performance comparison among different methods. The AUPR of protein-centric on varying sequence identity groups and the distributions of AUPR of term-centric on varying IC across MF, BP, and CC categories are in (a) and (b).}
 \label{fig:three_main_figures}
\end{figure*}

\subsection{PhaGO improves annotation of proteins by utilizing the contextual information}

In this section, we designed two experiments to evaluate how contextual proteins impact function prediction. In the first experiment, we compare two different usages of the protein embeddings from the foundation model: 1) using the per-residue embedding of a single protein as input, and 2) using joint embeddings of multiple proteins with genomic context. A model named `Trans' is designed for a single protein input, which uses the amino acids as tokens and utilizes the Transformer to learn the relationship of residues. A detailed description of the methods is in the Supplementary File. For the multiple protein input, the PhaGO is used to learn the protein associations and predict the GO terms. In the second experiment, we compare the performance of PhaGO in different protein context sizes by gradually increasing the number of context proteins. This step-by-step analysis provides insights into how the augmentation of context information influences the model's performance.


Fig. \ref{fig:context} shows the results for the first experiment. Based on the ESM2-12 model, a comparison between $\mathrm{Trans}_{\mathrm{BASE}}$  and $\mathrm{PhaGO}_{\mathrm{BASE}}$ reveals that BP and CC exhibit improvements of 7.3\% and 3.8\% in AUPR, respectively. Additionally, Fmax shows enhancements of 3.52\% and 2.23\% for BP and CC, respectively. Similarly, based on the ESM2-33 model, a comparison between $\mathrm{Trans}_{\mathrm{LARGE}}$  and $\mathrm{PhaGO}_{\mathrm{LARGE}}$  indicates that MF demonstrates the most significant improvement, with increases of 3.90\% and 2.75\% in AUPR and Fmax, respectively.

In the second experiment, we fed sentences with an increasing number of proteins into our model to show the impact of different contextual information. This involves creating three datasets:

\begin{itemize}
\item \textbf{Length\textgreater{}2 dataset}. We select protein sentences whose length is two or greater. Our goal is to preserve the original context information for our subsequent annotation.

\item \textbf{Length=1 dataset}. From the `Length\textgreater{}2' dataset, we extract individual proteins by dividing the selected protein sentences. This dataset ensures that each sentence includes only one protein, thereby removing the contextual information.
\item \textbf{Length=2 dataset}. Taking the protein sentences from the `Length\textgreater{}2' dataset, we divide them into pairs of two proteins with one overlap. Each sentence contains two proteins, representing an increase of one context protein compared to the `Length=1' dataset.
\end{itemize}

By inputting three datasets containing varying levels of contextual information into our model, we observed notable trends in performance, as illustrated in Figure \ref{fig:context_length}. The results indicate a consistent pattern of performance enhancement as the number of context proteins is progressively augmented. As more contextual information is provided, the model better understands the relationships and interactions between proteins, resulting in improved predictions of protein functions. These findings emphasize the importance of considering contextual proteins and their impact on protein function prediction tasks.

\begin{figure*}[t] 
  \centering
  \includegraphics[width=\linewidth]{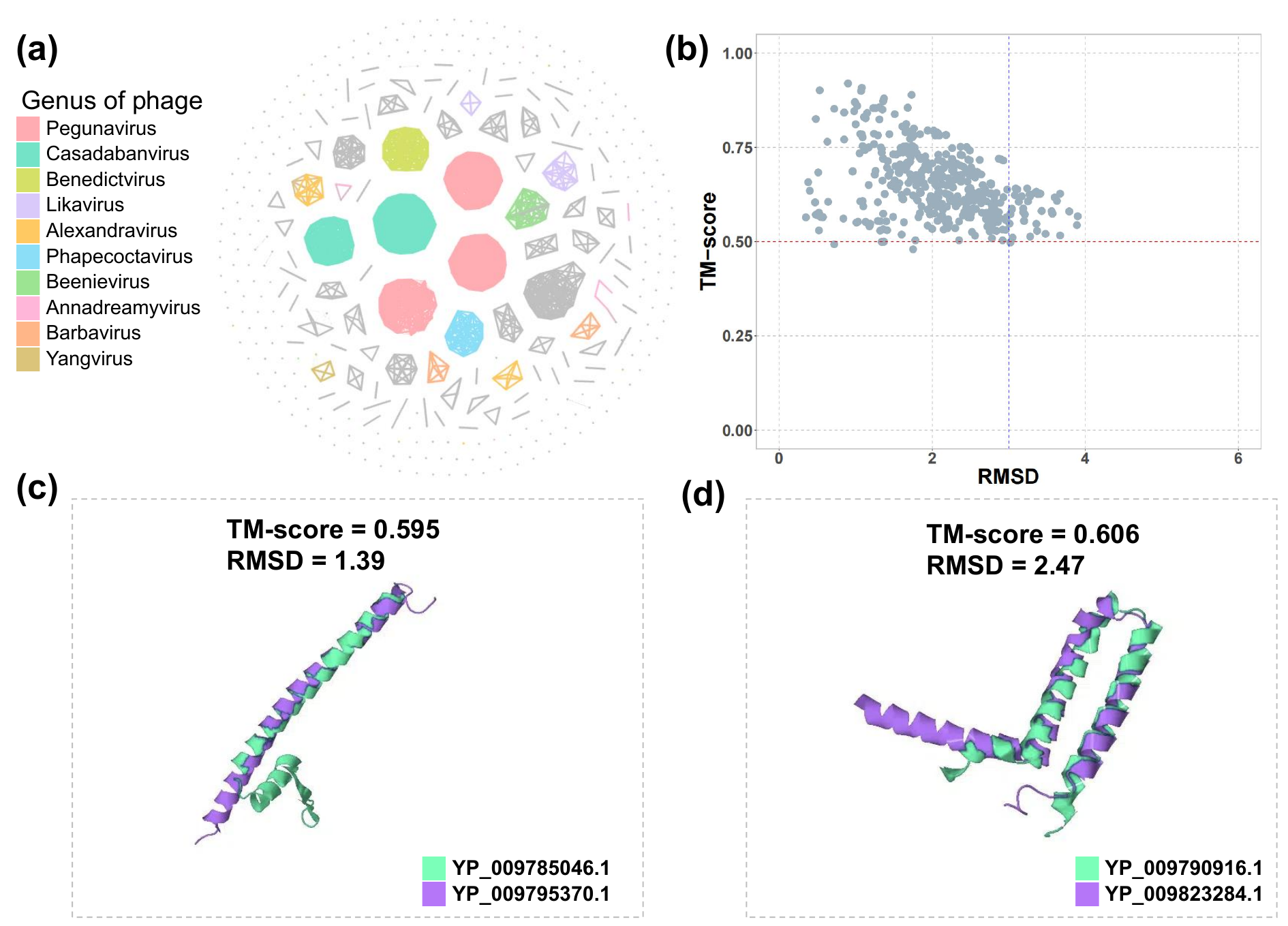}
  \caption{Distribution of holin proteins among the top 10 phage genera (a) and their structure similarity with the known holin proteins in  (b). (c) and (d) show the 3D structures of two identified holin proteins (YP\_009795370.1 and YP\_009823284.1) and known holin proteins in the database (YP\_009785046.1 and YP\_009790916.1). The TM-score and RMSD are two metrics to quantify the structure similarity of proteins. TM-score larger than 0.5 or RMSD less than 3 Å mean highly similar.}
 \label{fig:holin_proteins}
\end{figure*}

\subsection{PhaGO shows superior performance in annotating remote homologous proteins}

In this section, we evaluate PhaGO's predictive capability with different levels of sequence identity. The test dataset was partitioned into three distinct groups based on alignment with the training data: `no-alignment' for proteins lacking alignment, `min-40\%' for those below 40\% identity, and `40\%-100\%' for proteins with 40-100\% identity. These categories represent proteins with no similarity, low similarity, and moderate to high similarity to the training set, respectively. As shown in Fig. \ref{fig:three_main_figures}(b), the AUPR of all methods improved with increased sequence identity for all three GO categories. For the high-similarity dataset, the alignment-based method exhibits excellent performance, and $\mathrm{PhaGO}^+$ demonstrates comparable results in three ontologies. It suggests that both methods can effectively predict protein functions when the dataset aligns well with the training dataset. However, for the dataset that has no alignment with the training dataset, $\mathrm{PhaGO}^+$ stands out with impressive AUPR. Specifically, $\mathrm{PhaGO}^+$ achieves AUPR values of 0.7524, 0.8478, and 0.8210 for the BP, CC, and MF. These values represent improvements of 5.68\%, 6.78\%, and 5.75\% compared to the performance of the second-best method. The term-centric results are shown with a similar trend in supplementary Fig. 1(a). Additionally, the percentage of no-alignment proteins accounts for 27.93\%, 55.70\%, and 27.62\% of the test dataset for BP, CC, and MF, respectively. These results highlight the robustness and effectiveness of $\mathrm{PhaGO}^+$ in predicting protein functions, especially for low-similarity proteins. 

We continue to analyze the impact of contextual protein information on different level-similarity groups. The results are shown in the Supplementary  Fig. 1(b). Focusing on the no-alignments dataset, both $\mathrm{PhaGO}_{\mathrm{BASE}}$ and $\mathrm{PhaGO}_{\mathrm{LARGE}}$ demonstrate improvements compared to their respective counterparts. On one hand, $\mathrm{PhaGO}_{\mathrm{BASE}}$ shows performance gains of 10.18\%, 6.5\%, and 1.11\% for BP, CC, and MF categories, respectively. On the other hand, $\mathrm{PhaGO}_{\mathrm{LARGE}}$ exhibits improvements of 5.33\%, 3.87\%, and 7.91\% for BP, CC, and MF categories, respectively. 

\subsection{PhaGO enhances annotation with focus on minority class GO terms}

To examine the ability of PhaGO on the GO terms of different popularities, we split GO terms into three groups based on the information content (IC) of GO shown in Eqn. \ref{Eq:IC}. ${f}\left(l\right)$ is the frequency of the GO term $l$ in the training dataset. Higher IC values mean fewer proteins annotated by the GO term labels.

\begin{equation}
\mathrm{IC}\left(l\right)=-\log _2 f\left(l\right).
\label{Eq:IC}
\end{equation}



The experiment results in Fig. \ref{fig:three_main_figures}(b) demonstrate that all methods consistently performed well in the majority labels of GO terms. However, $\mathrm{PhaGO}^+$ demonstrates a distinct advantage in predicting minority GO terms, surpassing the other methods and achieving the highest performance across all three ontologies. Specifically, $\mathrm{PhaGO}^+$ achieves medium AUPR of 0.8801, 0.9043, and 0.8105 for BP, CC, and MF in the smallest GO terms group, respectively. This indicates that even for infrequently occurring GO terms, $\mathrm{PhaGO}^+$ can make an accurate prediction. 

We also further investigate the impact of the context proteins on the different GO terms. The results are shown in Supplementary Fig. 1(c). Focusing on the smallest GO terms group, both $\mathrm{PhaGO}_{\mathrm{BASE}}$  and $\mathrm{PhaGO}_{\mathrm{LARGE}}$ demonstrate improvements in performance. For the BP and CC ontologies, $\mathrm{PhaGO}_{\mathrm{BASE}}$ shows performance gains of 4.5\% and 3.55\% in AUPR, respectively. Moreover, it achieves comparable results for MF. In addition, $\mathrm{PhaGO}_{\mathrm{LARGE}}$ exhibits improvements of 3.71\%, 2.29\%, and 2.80\% for BP, CC, and MF categories, respectively. These results highlight the benefits of incorporating context proteins in predicting fewer GO terms.

\subsection{PhaGO enables protein function annotation without relying on homology search}

To showcase the utility of PhaGO in annotating proteins that lack homology search results, we explore its application in the analysis of phage's holin proteins. The holin protein is a small membrane protein that plays a crucial role in lysing bacterial hosts by triggering the formation of pores that disrupt the host cell membrane \cite{holin_proteins}. It controls the release of phages and the completion of the lytic cycle, underscoring the significance of the intricate interplay between phages and their host organisms. However, according to the protein annotation of phages in the RefSeq database, over $448$ genera have no annotated holin proteins, indicating that holin proteins may be very diverse across different phages. In this experiment, we apply PhaGO to annotate possible holin proteins.

According to statistical analysis of GO terms for the well-studied holin proteins from UniproKB, we manually selected six GO terms as their indicator. The details of selecting GO terms are shown in the Supplementary file. We input all proteins from $448$ genera into $\mathrm{PhaGO}^+$ and identified 688 potential holin proteins spanning $262$ genera. After identifying possible holins, we clustered them to analyze their relationship. To accomplish this, we aligned them all against all and selected alignment with identity and coverage larger than 90. Gephi was used to represent the relationships among proteins visually. The results depicting the top 10 phage genera in Fig. \ref{fig:holin_proteins}(a). The genera of phage are from the RefSeq annotations. An evident observation is the high conservation of holin proteins within the same genus, mirroring a common pattern observed among known holin proteins in phage genomes. 

In addition, we aligned them with the known holin proteins using BLASTP \citep{altschul1997gapped} with e-value 1e-5. 590 proteins have no alignment, indicating the high diversity of holin proteins. Then, we searched the annotation of 688 proteins from the UniProtKB database. Out of these, 335 proteins are recommended for annotation as holin, 171 proteins are labeled as uncharacterized proteins, and 87 proteins are categorized as membrane proteins. These top three annotations collectively account for 86.2\% of the total proteins.

To further examine the identified holin proteins without alignment, we employed ESMFold \citep{esmfold} to predict their three-dimensional (3D) structures, which are very fast and can get comparable predictions with AlphaFold \citep{jumper2021highly}. We found that despite having low sequence similarity, 590 identified holin proteins exhibit structural homology with the known holin proteins. The result is shown in \ref{fig:holin_proteins}(b). The TM-score and the Root Mean Square Deviation (RMSD) value are calculated by the TM-align tool \citep{zhang2005tm}. Fig. \ref{fig:holin_proteins}(c) and (d) are visualizations of the two putative holin proteins identified by our tool. In conclusion, the experiments provide further evidence of the great potential of PhaGO as a valuable tool for viral protein annotation. In addition, the information and 3D structure of the 688 holins are available in the Supplementary data.

\section{Conclusion and Discussion}
In this work, we proposed a method named PhaGO/$\mathrm{PhaGO}^+$ for protein function annotation of phages. The major improvement in our approach can be attributed to utilizing the properties of phages and the foundation model. The Transformer model is used to learn the relationship of the genomic context proteins. Our experiments compared four methods including alignment-based and deep learning-based. They have shown that PhaGO can achieve the highest AUPR and $F_{max}$ across all three ontologies, especially on low-similarity and minority GO term labels. Furthermore, we investigated the impact of incorporating context proteins into the annotation process and observed that PhaGO exhibits significant improvements compared to using only individual proteins as input. Notably, PhaGO plays a crucial role in enabling the characterization of unannotated proteins, making it a valuable tool for biological discovery and in-depth investigations. 

However, the major challenge for better viral protein annotation is the limited proteins with annotations, which limits the performance of PhaGO and general learning-based tools. By including proteins with enriched GO terms, we can augment the pool of context proteins available for analysis, thereby significantly enhancing the performance of PhaGO. Furthermore, the incorporation of additional proteins will amplify the number of GO term labels and facilitate the annotation of phage proteins at a more specific and detailed level. This advancement in annotation granularity will provide valuable insights into the intricate functional characteristics of these proteins. In the future, incorporating additional features such as structure information derived from GO graphs and textual descriptions of proteins is a valuable direction for further improving the annotation process. 

\section{Data availability}
PhaGO is implemented in Python, which can be downloaded
from \href{https://github.com/jiaojiaoguan/PhaGO}{https://github.com/jiaojiaoguan/PhaGO}.

\section{SUPPLEMENTARY DATA}
Supplementary Data are available Online.

\section{Competing interests}
No competing interest is declared.



\section{Acknowledgments}
The authors thank the anonymous reviewers for their valuable suggestions. This work is supported by GRF 11209823, City University of Hong Kong 7005866, and ARG 9667256.

\bibliographystyle{unsrtnat}

\begin{thebibliography}{10}

\bibitem{uniprot2023uniprot}
Uni{P}rot: the universal protein knowledgebase in 2023.
\newblock {\em Nucleic Acids Research}, 51(D1):D523--D531, 2023.

\bibitem{abdelrahman2021phage}
Fatma Abdelrahman, Maheswaran Easwaran, Oluwasegun~I Daramola, Samar Ragab, Stephanie Lynch, Tolulope~J Oduselu, Fazal~Mehmood Khan, Akomolafe Ayobami, Fazal Adnan, Eduard Torrents, et~al.
\newblock Phage-encoded endolysins.
\newblock {\em Antibiotics}, 10(2):124, 2021.

\bibitem{altschul1997gapped}
Stephen~F Altschul, Thomas~L Madden, Alejandro~A Sch{\"a}ffer, Jinghui Zhang, Zheng Zhang, Webb Miller, and David~J Lipman.
\newblock Gapped {BLAST} and {PSI-BLAST}: a new generation of protein database search programs.
\newblock {\em Nucleic Acids Research}, 25(17):3389--3402, 1997.

\bibitem{azimi2019phage}
Taher Azimi, Mehrdad Mosadegh, Mohammad~Javad Nasiri, Sahar Sabour, Samira Karimaei, and Ahmad Nasser.
\newblock Phage therapy as a renewed therapeutic approach to mycobacterial infections: a comprehensive review.
\newblock {\em Infection and Drug Resistance}, pages 2943--2959, 2019.

\bibitem{buchfink2015fast}
Benjamin Buchfink, Chao Xie, and Daniel~H Huson.
\newblock Fast and sensitive protein alignment using {DIAMOND}.
\newblock {\em Nature methods}, 12(1):59--60, 2015.

\bibitem{cao2021tale}
Yue Cao and Yang Shen.
\newblock {TALE}: {T}ransformer-based protein function {A}nnotation with joint sequence--{L}abel {E}mbedding.
\newblock {\em Bioinformatics}, 37(18):2825--2833, 2021.

\bibitem{cobian2016viruses}
Ana~Georgina Cobi{\'a}n~G{\"u}emes, Merry Youle, Vito~Adrian Cant{\'u}, Ben Felts, James Nulton, and Forest Rohwer.
\newblock Viruses as winners in the game of life.
\newblock {\em Annual review of virology}, 3(1):197--214, 2016.

\bibitem{gene2019gene}
Gene~Ontology Consortium.
\newblock The gene ontology resource: 20 years and still {GO}ing strong.
\newblock {\em Nucleic Acids Research}, 47(D1):D330--D338, 2019.

\bibitem{diaz2014bacteria}
Samuel~L D{\'\i}az-Mu{\~n}oz and Britt Koskella.
\newblock Bacteria--phage interactions in natural environments.
\newblock {\em Advances in applied microbiology}, 89:135--183, 2014.

\bibitem{edera2022anc2vec}
Alejandro~A Edera, Diego~H Milone, and Georgina Stegmayer.
\newblock Anc2vec: embedding gene ontology terms by preserving ancestors relationships.
\newblock {\em Briefings in Bioinformatics}, 23(2):bbac003, 2022.

\bibitem{elnaggar2007prottrans}
Ahmed Elnaggar, Michael Heinzinger, Christian Dallago, Ghalia Rihawi, Yu~Wang, Llion Jones, Tom Gibbs, Tamas Feher, Christoph Angerer, Martin Steinegger, et~al.
\newblock Prot{T}rans: {T}owards cracking the language of life’s code through self-supervised deep learning and high performance computing. arxiv 2020.
\newblock {\em arXiv preprint arXiv:2007.06225}, 2007.

\bibitem{fernandez2018phage}
Luc{\'\i}a Fern{\'a}ndez, Ana Rodr{\'\i}guez, and Pilar Garc{\'\i}a.
\newblock Phage or foe: an insight into the impact of viral predation on microbial communities.
\newblock {\em The ISME journal}, 12(5):1171--1179, 2018.

\bibitem{jumper2021highly}
John Jumper, Richard Evans, Alexander Pritzel, Tim Green, Michael Figurnov, Olaf Ronneberger, Kathryn Tunyasuvunakool, Russ Bates, Augustin {\v{Z}}{\'\i}dek, Anna Potapenko, et~al.
\newblock Highly accurate protein structure prediction with {A}lpha{F}old.
\newblock {\em nature}, 596(7873):583--589, 2021.

\bibitem{DeepGOplus}
Maxat Kulmanov and Robert Hoehndorf.
\newblock {Deep{GOP}lus: improved protein function prediction from sequence}.
\newblock {\em Bioinformatics}, 36(2):422--429, 07 2019.

\bibitem{esm2}
Zeming Lin, Halil Akin, Roshan Rao, Brian Hie, Zhongkai Zhu, Wenting Lu, Nikita Smetanin, Robert Verkuil, Ori Kabeli, Yaniv Shmueli, Allan dos Santos~Costa, Maryam Fazel-Zarandi, Tom Sercu, Salvatore Candido, and Alexander Rives.
\newblock Evolutionary-scale prediction of atomic-level protein structure with a language model.
\newblock {\em Science}, 379(6637):1123--1130, 2023.

\bibitem{esmfold}
Zeming Lin, Halil Akin, Roshan Rao, Brian Hie, Zhongkai Zhu, Wenting Lu, Nikita Smetanin, Robert Verkuil, Ori Kabeli, Yaniv Shmueli, Allan dos Santos~Costa, Maryam Fazel-Zarandi, Tom Sercu, Salvatore Candido, and Alexander Rives.
\newblock Evolutionary-scale prediction of atomic-level protein structure with a language model.
\newblock {\em Science}, 379(6637):1123--1130, 2023.

\bibitem{lin2023evolutionary}
Zeming Lin, Halil Akin, Roshan Rao, Brian Hie, Zhongkai Zhu, Wenting Lu, Nikita Smetanin, Robert Verkuil, Ori Kabeli, Yaniv Shmueli, et~al.
\newblock Evolutionary-scale prediction of atomic-level protein structure with a language model.
\newblock {\em Science}, 379(6637):1123--1130, 2023.

\bibitem{Bacteriophage_therapy2}
Hao Ling, Xinyu Lou, Qiuhua Luo, Zhonggui He, Mengchi Sun, and Jin Sun.
\newblock Recent advances in bacteriophage-based therapeutics: {I}nsight into the post-antibiotic era.
\newblock {\em Acta Pharmaceutica Sinica B}, 12(12):4348--4364, 2022.

\bibitem{Bacteriophage_therapy}
Mikeljon~P Nikolich and Andrey~A Filippov.
\newblock Bacteriophage therapy: {D}evelopments and directions.
\newblock {\em Antibiotics}, 9(3):135, 2020.

\bibitem{outeiral2024codon}
Carlos Outeiral and Charlotte~M Deane.
\newblock Codon language embeddings provide strong signals for use in protein engineering.
\newblock {\em Nature Machine Intelligence}, 6(2):170--179, 2024.

\bibitem{pan2023pfresgo}
Tong Pan, Chen Li, Yue Bi, Zhikang Wang, Robin~B Gasser, Anthony~W Purcell, Tatsuya Akutsu, Geoffrey~I Webb, Seiya Imoto, and Jiangning Song.
\newblock P{F}res{GO}: an attention mechanism-based deep-learning approach for protein annotation by integrating gene ontology inter-relationships.
\newblock {\em Bioinformatics}, 39(3):btad094, 2023.

\bibitem{radivojac2013large}
Predrag Radivojac, Wyatt~T Clark, Tal~Ronnen Oron, Alexandra~M Schnoes, Tobias Wittkop, Artem Sokolov, Kiley Graim, Christopher Funk, Karin Verspoor, Asa Ben-Hur, et~al.
\newblock A large-scale evaluation of computational protein function prediction.
\newblock {\em Nature Methods}, 10(3):221--227, 2013.

\bibitem{holin_proteins}
Erlan Ramanculov and Ry~Young.
\newblock Genetic analysis of the {T}4 holin: timing and topology.
\newblock {\em Gene}, 265(1):25--36, 2001.

\bibitem{rives2021biological}
Alexander Rives, Joshua Meier, Tom Sercu, Siddharth Goyal, Zeming Lin, Jason Liu, Demi Guo, Myle Ott, C~Lawrence Zitnick, Jerry Ma, et~al.
\newblock Biological structure and function emerge from scaling unsupervised learning to 250 million protein sequences.
\newblock {\em Proceedings of the National Academy of Sciences}, 118(15):e2016239118, 2021.

\bibitem{santos2018exploiting}
S{\'\i}lvio~B Santos, Ana~Rita Costa, Carla Carvalho, Franklin~L N{\'o}brega, and Joana Azeredo.
\newblock Exploiting bacteriophage proteomes: the hidden biotechnological potential.
\newblock {\em Trends in biotechnology}, 36(9):966--984, 2018.

\bibitem{shibayama2011phage}
Youtaro Shibayama and Eric~R Dabbs.
\newblock Phage as a source of antibacterial genes: {M}ultiple inhibitory products encoded by {R}hodococcus phage {YF}1.
\newblock {\em Bacteriophage}, 1(4):195--197, 2011.

\bibitem{song2016identification}
Jun Song, Feifei Xia, Haiyan Jiang, Xinwei Li, Liyuan Hu, Pengjuan Gong, Liancheng Lei, Xin Feng, Changjiang Sun, Jingmin Gu, et~al.
\newblock Identification and characterization of holgh15: the holin of staphylococcus aureus bacteriophage gh15.
\newblock {\em Journal of General Virology}, 97(5):1272--1281, 2016.

\bibitem{steinegger2019hh}
Martin Steinegger, Markus Meier, Milot Mirdita, Harald V{\"o}hringer, Stephan~J Haunsberger, and Johannes S{\"o}ding.
\newblock {HH}-suite3 for fast remote homology detection and deep protein annotation.
\newblock {\em BMC bioinformatics}, 20:1--15, 2019.

\bibitem{terzian2021phrog}
Paul Terzian, Eric Olo~Ndela, Clovis Galiez, Julien Lossouarn, Rub{\'e}n~Enrique P{\'e}rez~Bucio, Robin Mom, Ariane Toussaint, Marie-Agn{\`e}s Petit, and Fran{\c{c}}ois Enault.
\newblock {PHROG}: families of prokaryotic virus proteins clustered using remote homology.
\newblock {\em NAR Genomics and Bioinformatics}, 3(3):lqab067, 2021.

\bibitem{totte2017successful}
Joan~EE Tott{\'e}, Martijn~B van Doorn, and Suzanne~GMA Pasmans.
\newblock Successful treatment of chronic {S}taphylococcus aureus-related dermatoses with the topical endolysin {S}taphefekt sa. 100: a report of 3 cases.
\newblock {\em Case reports in dermatology}, 9(2):19--25, 2017.

\bibitem{mcl}
Stijn Van~Dongen.
\newblock Graph {C}lustering via a {D}iscrete {U}ncoupling {P}rocess.
\newblock {\em SIAM Journal on Matrix Analysis and Applications}, 30(1):121--141, 2008.

\bibitem{vaswani2017attention}
Ashish Vaswani, Noam Shazeer, Niki Parmar, Jakob Uszkoreit, Llion Jones, Aidan~N Gomez, {\L}ukasz Kaiser, and Illia Polosukhin.
\newblock Attention is all you need.
\newblock {\em Advances in neural information processing systems}, 30, 2017.

\bibitem{WANG2024120859}
Donglin Wang, Jiayu Shang, Hui Lin, Jinsong Liang, Chenchen Wang, Yanni Sun, Yaohui Bai, and Jiuhui Qu.
\newblock Identifying {ARG}-carrying bacteriophages in a lake replenished by reclaimed water using deep learning techniques.
\newblock {\em Water Research}, 248:120859, 2024.

\bibitem{tailspike}
Yiyan Yang, Keith Dufault-Thompson, Wei Yan, Tian Cai, Lei Xie, and Xiaofang Jiang.
\newblock {Large-scale genomic survey with deep learning-based method reveals strain-level phage specificity determinants}.
\newblock {\em GigaScience}, 13:giae017, 04 2024.

\bibitem{zeng2024metagenomic}
Shuqin Zeng, Alexandre Almeida, Shiping Li, Junjie Ying, Hua Wang, Yi~Qu, R~Paul~Ross, Catherine Stanton, Zhemin Zhou, Xiaoyu Niu, et~al.
\newblock A metagenomic catalog of the early-life human gut virome.
\newblock {\em Nature Communications}, 15(1):1864, 2024.

\bibitem{zhang2005tm}
Yang Zhang and Jeffrey Skolnick.
\newblock {TM}-align: a protein structure alignment algorithm based on the tm-score.
\newblock {\em Nucleic Acids Research}, 33(7):2302--2309, 2005.

\bibitem{zhu2022integrating}
Yi-Heng Zhu, Chengxin Zhang, Dong-Jun Yu, and Yang Zhang.
\newblock Integrating unsupervised language model with triplet neural networks for protein gene ontology prediction.
\newblock {\em PLOS Computational Biology}, 18(12):e1010793, 2022.

\end{thebibliography}


\end{document}